\documentclass[12pt]{article}
\begin{document}
\title{Anisotropic Bose-Einstein Condensates and Completely Integrable Dynamical Systems}
\author{F.~Haas  \\
Laborat\'orio Nacional de Computa\c{c}\~ao Cient\'{\i}fica \\ Departamento de
Matem\'atica Aplicada e Computacional \\
Av. Get\'ulio Vargas, 333 \\
25651-070, Petr\'opolis, RJ - Brazil\\
ferhaas@lncc.br}
\date{\today}

\maketitle

\begin{abstract} \noindent
A Gaussian {\it ansatz} for the wave function of two-dimensional
harmonically trapped anisotropic Bose-Einstein condensates is
shown to lead, via a variational procedure, to a coupled system of
two second-order, nonlinear ordinary differential equations. This
dynamical system is shown to be in the general class of Ermakov
systems. Complete integrability of the resulting Ermakov system is
proven. Using the exact solution, collapse of the condensate is
analyzed in detail. Time-dependence of the trapping potential is
allowed.
\end{abstract}

{\it PACS numbers: 03.75.Fi, 05.45.-a, 32.80.Pj}

\section{Introduction}

We consider the mean-field theory for two-di\-men\-sio\-nal interacting Bose-Einstein
condensates, described by the Gross-Pitaevskii equation \cite{becreview}. We suppose a
time-dependent harmonic trap. Introducing a Gaussian {\it ansatz} for the amplitude of
the wave-function of the Bose-Einstein condensate, there results a coupled nonlinear
system of second-order ordinary differential equations for the time evolution of the
system \cite{Rybin}-\cite{Dalfovo}. Usually, this system is studied only in the
isotropic case for which the condensate has equal width in all directions
\cite{Rybin}. The purpose of this work is to show that in the case of two spatial
dimensions the isotropy assumption is not needed, since the nonlinear system for the
Bose-Einstein condensate can be cast in the form of a completely integrable Lagrangian
Ermakov system \cite{Cervero, HG1, HG2}. Moreover, using the formalism of Lagrangian
Ermakov systems, we are able to extract detailed information about the time evolution
of the condensate. These results have an increased importance in view of the recent
experimental achievement of Bose-Einstein condensation in quasi-one and quasi-two
dimensions \cite{Gorlitz}-\cite{Morsch}. For instance, the present theory provides an
accurate estimate for the critical time of collapse of the condensate (see equation
(\ref{x3}) bellow), when collapse can occurs.

Ermakov systems \cite{Ermakov} have attracted interest in the last three decades or so
due to both their physical applications and nice mathematical properties . The central
feature of Ermakov systems is the existence of a constant of motion, the so called
Ermakov invariant \cite{RR1}, which generalizes the Lewis invariant for the
time-dependent harmonic oscillator \cite{L67}. The Ermakov invariant allows the
construction of a nonlinear superposition law \cite{RR83} between the solutions of the
differential equations of the Ermakov system. Moreover, the existence of the Ermakov
invariant allows to linearize \cite{Ath,HG} the Ermakov system in many cases. Other
theoretical studies on Ermakov systems concerns its Lie symmetry structure
\cite{Leach, HGO}, the existence of additional constants of motion \cite{GO} and the
extension of the Ermakov systems concept to higher dimensions \cite{Leach}. From the
physical point of view, Ermakov systems have found applications in several problems,
such as cosmological particle creation \cite{R79}, nonlinear optics \cite{Goncha1,
Goncha2} and propagation of shallow water waves \cite{coupinney}.

As will be seen in the following, two-di\-men\-sio\-nal Bose-Einstein condensates can
be described by a nonlinear dynamical system of Ermakov type, admitting a Lagrangian
formulation.  From the Ermakov property, one constant of motion, the Ermakov
invariant, is immediately derived. The existence of a variational principle suggests
the use of symmetry tools for the search of additional invariants. Here, we look for
Noether point symmetries, which preserves the action functional up to addition of a
numerical constant. We show that the Lagrangian Ermakov system for
two-di\-men\-sio\-nal time-dependent Bose-Einstein condensates do indeed possess
Noether point symmetry. The associated Noether constant of motion and the Ermakov
invariant can then be used to construct the exact solution for the equations of
motion.

We remark that Bose-Einstein condensation cannot occur in uniform
two-di\-men\-sio\-nal  and one-di\-men\-sio\-nal  gases at finite temperature, due to
destabilizing thermal effects. However, we are not faced with this problem here since
we treat non-uniform condensates under the influence of harmonic traps. Also, we
stress that completely integrable Lagrangian Ermakov systems of similar character also
arises in the study of the propagation of elliptic Gaussian beams in nonlinear optics
\cite{Goncha1, Goncha2}. These problems in nonlinear optics are also described by a
nonlinear Sch\"odinger equation formally identical to the Gross-Pitaevskii equation.
However, the distinctive feature of the present work is the presence of explicitly
time-dependent harmonic traps, as far as we know a feature not previously considered
in the literature \cite{Goncha1, Goncha2}. As will be seen in Section 4, there are
concrete examples in which the details of the time-dependence of the harmonic fields
can decisively influence on such relevant questions such as the critical time for
collapse of the condensate, when collapse is in order.

The work is organized as follows. In Section 2, we describe the basic steps for the
conversion of the problem of solving the Gross-Pitaevskii equation in two spatial
dimensions to the problem of solving a coupled, nonlinear system of second-order
ordinary differential equations. This reduction can occur in view of the Gaussian {\it
ansatz} for the condensate wave function. Thanks to the harmonic character of the
trap, we formulate the reduced system as an Lagrangian Ermakov system. In Section 3,
we use the Lagrangian formalism to derive an additional invariant by means of
Noether's theorem. In Section 4 we present the main physical consequences which follow
from this methodology. In Section 5, the joint conservation of the Ermakov invariant
and the Noether invariant allows to obtain the exact solution of the system in terms
of a particular solution for an auxiliary equation whose form is to be considered in
the following. The whole formalism is exemplified in Section 6 for some particular
choices of time-dependent frequencies of the harmonic trap. In Section 7 we present
our final remarks and conclusion.

\section{Bo\-se-Eins\-tein Con\-den\-sa\-tes as Er\-ma\-kov Sys\-tems}

Our starting point is the Gross-Pitaevskii equation \cite{becreview},
\begin{equation}
\label{eq1} - \frac{\hbar^2}{2m}\nabla^{2}\psi + g|\psi|^{2}\psi + U\psi =
i\hbar\frac{\partial\psi}{\partial t} \,,
\end{equation}
which gives a mean-field description for Bose-Einstein condensates. Here, $\psi =
\psi({\bf x},t)$ is the condensate wave function, while $g$ is the coupling constant.
In addition, $m$ is the mass of the particles of the condensate and $U = U({\bf x},t)$
is a confining potential. More specifically, we are looking for the
two-di\-men\-sio\-nal case so that $\psi = \psi(x,y,t)$, $U = U(x,y,t)$.

In what follows we will consider time-dependent harmonic traps. More specifically, we
put
\begin{equation}
U = \frac{m}{2}\omega_{\bot}^{2}(t)(x^2 + y^2) \,.
\end{equation}

The two-di\-men\-sio\-nal  Gross-Pitaevskii equation can be derived \cite{Salasnich}
from the variational principle $\delta\,S = 0$, with
\begin{equation}
S = \int dt d{\bf x}\psi^{*}\left(i\hbar\frac{\partial\psi}{\partial t} +
\frac{\hbar^{2}\nabla^{2}}{2m} - U - \frac{g}{2}|\psi|^2\right)\psi \,,
\end{equation}
where $d{\bf x} = dx dy$.

Following references \cite{Rybin}-\cite{Dalfovo}, we introduce a Gaussian wave
function,
\begin{equation}
\label{eq7} \psi =
\left(\frac{N}{\pi\alpha_{1}\alpha_{2}}\right)^{1/2} \exp(-
\frac{x^{2}}{2\alpha_{1}^2} -
\frac{y^{2}}{2\alpha_{2}^2})\exp(i(\beta_{1}x^2 + \beta_{2}y^2))
\,,
\end{equation}
where $\alpha_i$ and $\beta_i$ are real functions, depending only on time, and $N$ is
the total number of particles. The trial wave function is normalized as
\begin{equation}
\int |\psi|^2 = N \,.
\end{equation}
The Gaussian {\it ansatz} is a reasonable proposition for weak coupling
since the ground state of the simple
harmonic oscillator is a Gaussian wave function. Notice, however, that we are
admitting a time-dependent frequency $\omega_{\bot}(t)$ here.

After in\-ser\-ting the Gaus\-sian ansatz on the two-di\-men\-sio\-nal
Gross-Pi\-taevskii action, the result is
\begin{equation}
S = \int dt {\cal L}({\bf\alpha},{\bf\beta},\dot{\bf\beta}) \,,
\end{equation}
for
\begin{eqnarray}
{\cal L} &=& - \frac{N}{2}\bigl[\hbar(\alpha_{1}^{2}\dot\beta_{1} +
\alpha_{2}^{2}\dot\beta_{2}) + \frac{2\hbar^{2}}{m}(\beta_{1}^{2}\alpha_{1}^{2} +
\beta_{2}^{2}\alpha_{2}^{2})  \nonumber \\ &+&
\frac{\hbar^{2}}{2m}(\frac{1}{\alpha_{1}^2} + \frac{1}{\alpha_{2}^2}) +
\frac{m\omega_{\bot}^{2}(t)}{2}(\alpha_{1}^2 + \alpha_{2}^2) +
\frac{gN}{2\pi\alpha_{1}\alpha_{2}}\bigr] \,.
\end{eqnarray}
Regarding ${\bf\alpha}$ and ${\bf\beta}$ as coordinates on configuration space, we
obtain the following Euler-Lagrange equations for the Lagrangian ${\cal L}$,
\begin{equation}
\label{phase}
\beta_1 = \frac{m\dot\alpha_1}{2\hbar\alpha_1} \,, \quad \beta_2 =
\frac{m\dot\alpha_2}{2\hbar\alpha_2} \,,
\end{equation}
and
\begin{eqnarray}
\label{g1} \ddot\alpha_{1} + \omega_{\bot}^{2}(t)\alpha_{1} &=&
\frac{\hbar^2}{m^{2}\alpha_{1}^{3}} + \frac{gN}{2\pi m\alpha_{1}^{2}\alpha_{2}} \,, \\
\label{g2} \ddot\alpha_2 + \omega_{\bot}^{2}(t)\alpha_{2} &=&
\frac{\hbar^2}{m^{2}\alpha_{2}^3} + \frac{gN}{2\pi\,m\alpha_{2}^{2}\alpha_{1}} \,.
\end{eqnarray}

Let us introduce some scalings. Let
\begin{eqnarray}
\omega_{\bot} &=& \omega_{0}\omega \,, \quad \bar{t} = \omega_{0}t
\,, \quad \bar{g} = \frac{Nm\omega_{0}^{2}g}{2\pi\hbar^{2}} \\
\bar{\alpha_i} &=& \alpha_{i}/a_{0} \,, \quad a_{0} =
(\frac{\hbar}{m\omega_0})^{1/2} \,.
\end{eqnarray}
Here, $\omega_0$ is a typical frequency, for instance the temporal average of
$\omega(t)$, and $a_0$ is the oscillator typical length. Also, the quantity $\bar{g}$
can be shown to correspond to the ratio of the mean potential energy to the mean
kinetic energy.

Dropping bars, the result is
\begin{eqnarray}
\label{eq11}
\ddot\alpha_1 + \omega^{2}(t)\alpha_{1} &=&
\frac{1}{\alpha_{2}\alpha_{1}^{2}}\,F(\alpha_{2}/\alpha_{1}) \,, \\
\label{eq12}
\ddot\alpha_2 + \omega^{2}(t)\alpha_{2} &=&
\frac{1}{\alpha_{1}\alpha_{2}^{2}}\,G(\alpha_{1}/\alpha_{2}) \,,
\end{eqnarray}
where the functions $F$ and $G$ are defined by
\begin{eqnarray}
\label{eq13}
F(\alpha_{2}/\alpha_{1}) &=& \alpha_{2}/\alpha_1 + g \,, \\
\label{eq14}
G(\alpha_{1}/\alpha_{2}) &=& \alpha_{1}/\alpha_{2} + g \,.
\end{eqnarray}

System (\ref{eq11}-\ref{eq12}) is in the class of Ermakov systems. Moreover,
(\ref{eq11}-\ref{eq12}) is an example of the still more specialized  class of
Lagrangian Ermakov systems. These two properties, of being an Ermakov system and of
admitting a Lagrangian formulation, are decisive to understand the dynamics of the
system, as will be seen in the forthcoming sections. Once the solution for the Ermakov
system is found, the phase functions $\beta_1$ and $\beta_2$ are derived from
(\ref{phase}), thus completely describing the time-evolution of the Gaussian
condensate wave function.

\section{Noether Point Symmetries and Invariants}

All dynamical systems of the Ermakov form (\ref{eq11}-\ref{eq12})
have the constant of motion
\begin{equation}
\label{eq15} I = \frac{1}{2}(\alpha_{1}\dot\alpha_{2} -
\alpha_{2}\dot\alpha_{1})^2 +
\int^{\alpha_{2}/\alpha_{1}}\,F(\lambda)d\lambda +
\int^{\alpha_{1}/\alpha_{2}}\,G(\lambda)d\lambda \,,
\end{equation}
the Ermakov invariant, whatever the functions $F$ and $G$. It can be easily proven
that $dI/dt = 0$ along trajectories. For our purposes, it is better to use $K = I + 1$
which for the choice ({\ref{eq13}-\ref{eq14}) is given by
\begin{eqnarray}
\label{eq16} K &=& \frac{1}{2}(r^{2}\dot\theta)^2 + V(\theta) \,, \\
\label{eqq16} V(\theta) &=& \frac{2}{\sin^{2}2\theta}(1 + g\sin{2\theta}) \,.
\end{eqnarray}
where $r = (\alpha_{1}^2 + \alpha_{2}^{2})^{1/2}$, $\tan\theta =
\alpha_{2}/\alpha_{1}$ are polar coordinates, which are the most convenient for the
analysis here.

A central question for any dynamical system concerns the existence of a sufficient
number of constants of motion for integrability. Hence, it is important to find
additional invariants independent of $K$. To address this question, we first remark
that equations (\ref{eq11}-\ref{eq12}) with $F$ and $G$ given by
(\ref{eq13}-\ref{eq14}) are the Euler-Lagrange equations for the Lagrangian
\begin{equation}
\label{eq17} L = \frac{1}{2}(\dot{r}^2 + r^{2}\dot{\theta}^{2}) -
\frac{1}{2}\omega^{2}(t)\,r^2 - \frac{V(\theta)}{r^{2}} \,.
\end{equation}
The Lagrangian structure suggest the use of variational techniques for the search of
extra constants of motion. Among these techniques, Noether's theorem \cite{Sarlet}
plays a distinctive role due to the physical appealing of the method.

In our work \cite{HG2}, we make an extensive analysis of the Noether point symmetries
for Lagrangian Ermakov systems. Referring to \cite{HG2} for the details, we derive the
following infinitesimal Noether point symmetry transformation,
\begin{equation}
\label{eq18} \tilde{r} = r + \varepsilon\rho\dot\rho\,r \,, \quad \tilde{\theta} =
\theta \,, \quad \tilde{t} = t + \varepsilon\rho^2 \,,
\end{equation}
where $\varepsilon$ is an infinitesimal parameter and $\rho$ is any particular
solution to Pinney's \cite{pinney} equation
\begin{equation}
\label{eq19}
\ddot\rho + \omega^{2}(t)\rho = \frac{c}{\rho^3} \,.
\end{equation}
In (\ref{eq19}), $c$ is an arbitrary numerical constant. Taking a positive $c$
prevents $\rho$ from changing sign, which is a convenient property for our purposes.
It suffices to put $c = 1$, a convention adopted in what follows. Also, just one
particular solution for Pinney's equation is sufficient for the application of the
theory. In the following, we always consider the particular solution selected by the
initial conditions $\rho(0) = 1$, $\dot\rho(0) = 0$.

Again by the results of \cite{HG2}, it can be shown that the Noether symmetry
(\ref{eq18}) is associated to the Noether invariant
\begin{equation}
\label{eq20} J = \frac{1}{2}(\rho\dot{r} - \dot\rho r)^2 + \frac{r^2}{2\rho^2} +
K\frac{\rho^2}{r^2} \,.
\end{equation}
It can be readily checked that $dJ/dt = 0$ along the trajectories of (\ref{eq11}-\ref{eq12}).

In reference \cite{HG2} it is also shown that the Ermakov invariant is associated to a
dynamical Noether symmetry,
\begin{equation}
\label{eq21}
\tilde{r} = r + \varepsilon\tau\dot{r} \,,\quad \tilde{\theta} = \theta +
\varepsilon(- r^{2}\dot\theta + \tau\dot\theta) \,,\quad \tilde{t} = t + \varepsilon\tau \,,
\end{equation}
where $\tau = \tau(r,\theta,\dot{r},\dot\theta,t)$ is an arbitrary function. It is
apparent that there is no choice of $\tau$ for which (\ref{eq21}) can be put in the form of a
point symmetry, independent of velocities.

The Noether invariant allows to quickly find the exact solution in terms of a
particular solution of Pinney's equation. Let the quasi-invariance transformation
\begin{equation}
\label{eq22} R = r/\rho \,, \quad T = \int_{0}^{t}d\lambda/\rho(\lambda)^2 \,.
\end{equation}
In the new coordinates $(R,T)$, the Noether invariant becomes the energy-like quantity
\begin{equation}
\label{eq23} J = \frac{1}{2}\left(\frac{dR}{dT}\right)^2 + W(R) \,,
\end{equation}
where
\begin{equation}
\label{eqq24} W(R) = \frac{R^2}{2} + \frac{K}{R^2} \,.
\end{equation}
We now proceed to the main physical consequences.  Integration of the system Is given
in Section 5.

\section{Physical Interpretation}

The Noether invariant is formally the energy function of a one-dimensional
time-independent singular oscillator, whose trajectories can be found by quadrature.
Referring to the potential $V(R)$, we know in advance that
\begin{equation}
\label{eqqq24} K \leq 0
\end{equation}
is a necessary and sufficient condition for $R(T)$ to attain the zero value at some
critical rescaled time $T = T^*$. Indeed, for $K > 0$, there is a repulsive force
preventing collapse. For $K = 0$, $V(R)$ is the potential for the simple harmonic
oscillator, which goes to the origin at some time. Finally, for $K < 0$, there is an
attractive force which constrain $R$ to be zero at some time.

The main physical consequence of our approach is the following. Examining the
quasi-invariance transformation (\ref{eq22}) and taking into account that $\rho(t)$
does not change sign (due to the properties of Pinney's equation), we immediately
conclude that (\ref{eqqq24}) is also a necessary and sufficient condition for $r(t)$
to attain the zero value at some critical time $t = t^*$. At this critical time, the
density of the gas becomes infinity, characterizing the collapse of the condensate. At
$t = t^*$, the Gross-Pitaevskii equation ceases to be a reasonable model, since too
great densities are clearly not acceptable in the framework of mean field theories.
The form of $t^*$ depends on the details of the frequency $\omega(t)$, but in all
cases the ultimate evolution of the system is dictated by the value of $K$, which is
essentially the Ermakov invariant for the problem.

Looking more carefully the invariant $K$ as given by (\ref{eq16}), we see that $1 +
g\sin{2\theta(0)} \leq 0$ is a necessary condition for $K \leq 0$. As $\alpha_{1}(0)$
and $\alpha_{2}(0)$ are positive by definition, we have $\sin{2\theta(0)} > 0$.
Therefore, we conclude that
\begin{equation}
\label{ggg} g \leq - \frac{1}{sin{2\theta(0)}}
\end{equation}
is a necessary but not sufficient condition for collapse. Indeed, $K$ is composed by a
positive definite part, related to the angular momentum of the condensate, and a term
which can be non-positive when (\ref{ggg}) is satisfied. Therefore, attracting
Bose-Einstein condensates with negative $g$ satisfying (\ref{ggg}) does not
necessarily collapse if the initial angular momentum is sufficiently high. This is a
distinctive feature from previous works \cite{Rybin} in which the isotropy condition
$\alpha_{1} = \alpha_{2}$ sets the angular momentum to zero. For the isotropic cases,
we also have $\sin{2\theta} = 1$ and $K = 2(1 + g)$, so that the condition for
collapse is simply $g \leq -1$. To the best of our knowledge, the fundamental role of
angular momentum in the collapsing or not of two-dimensional Bose-Einstein condensates
does not seem to be previously pointed out in the literature.

We now provide some details of the mathematical analysis. Our main result is equation
(\ref{x3}) bellow which gives the critical time $t^*$.

\section{Integration of the System}

As for any one-dimensional time-independent potential system, the trajectories in
rescaled variables can be obtained from the second integration of the energy-like form
(\ref{eq23}),
\begin{equation}
\label{xxx} \frac{1}{\sqrt{2}}\int^{R}\frac{d\lambda}{(J - W(\lambda))^{1/2}} = T +
k_1 \,,
\end{equation}
where $k_1$ is a numerical constant. For simplicity only, we set arbitrarily the initial
condition as $R(0) = 1$, $R'(0) = 0$. For this choice (\ref{xxx}) implies
\begin{equation}
\label{x1} R^2 = \cos^{2}T + 2K\sin^{2}T \,,
\end{equation}
showing in a more explicit way the role of the sign of $K$ in the existence or not of
collapse. Indeed, for $K > 0$, $R^2$ is a positive definite quantity for all $T$.

For non-positive $K$, the rescaled critical time for collapse is given, from $R(T*) =
0$ and (\ref{x1}), by
\begin{equation}
\label{x2} T^* = \arctan\left(\frac{1}{\sqrt{2|K|}}\right) \,.
\end{equation}
The physical critical time $t^*$ follows from (\ref{eq22}) and (\ref{x2}),
\begin{equation}
\label{x3} \int_{0}^{t^*}\frac{d\lambda}{\rho^{2}(\lambda)} =
\arctan\left(\frac{1}{\sqrt{2|K|}}\right) \,.
\end{equation}
The physical critical time strongly depends on the form of $\rho(t)$ solving Pinney's
equation, and then on the time-dependence of the frequency of the trap.

Expressing (\ref{x1}) in terms of physical coordinates, there results
\begin{equation}
\label{x4} r^2 = \left(\cos^{2}(\int_{0}^{t}\frac{d\lambda}{\rho^{2}(\lambda)}) +
2K\sin^{2}(\int_{0}^{t}\frac{d\lambda}{\rho^{2}(\lambda)})\right) \rho^{2}(t) \,,
\end{equation}
which furnishes the exact solution for the radial variable in terms of a particular
solution for the Pinney equation (\ref{eq19}). This particular solution may be
obtained numerically whenever necessary. The exact solution (\ref{x4}) is such that
$r(0) = 1$, $\dot{r}(0) = 0$.

So far only the time evolution of the radial variable was considered. In order to
obtain the angular variable, we integrate a second time the Ermakov invariant as given
by (\ref{eq16}),
\begin{equation} \label{x5}
\frac{1}{\sqrt{2}}\int^{\theta}\frac{d\lambda}{(K - V(\lambda))^{1/2}} =
\int^{t}\frac{d\lambda}{r^{2}(\lambda)} + k_2 \,,
\end{equation}
where $k_2$ is a numerical constant. The evaluation of the right-hand side of
(\ref{x5}) depends on the form of $r(t)$ and then on the form of $\rho(t)$. The
left-hand side can be evaluated in terms of elliptic integrals \cite{Goncha1}. In the
continuation, we are mainly concerned with the time-evolution of the radial variable.
However, notice that a consequence of (\ref{x5}) is that collapse, when it occurs,
take place with angular velocity going to infinity \cite{Goncha1}.

In the next section, we illustrate the theory for some specific forms of the
frequency.

\section{Examples}

\subsection{Constant Frequency}

In the case of constant frequency, we can set
\begin{equation}
\omega = 1
\end{equation}
in our non-dimensional variables. Then, the solution for Pinney's equation with $\rho(0) =
1$, $\dot\rho(0) = 0$ is
\begin{equation}
\rho = 1 \,.
\end{equation}
Hence, the quasi-invariance transformation (\ref{eq22}) becomes the identity and the
formulae (\ref{x1}-\ref{x2}) translate directly to physical variables. In other
words,
\begin{equation}
\label{xx1} r^2 = \cos^{2}t + 2K\sin^{2}t \,,
\end{equation}
and, for non-positive $K$, collapse take place at
\begin{equation}
\label{xx2} t^* = \arctan\left(\frac{1}{\sqrt{2|K|}}\right) \,.
\end{equation}
This critical time is a decaying function of $|K|$, as it should be. That is, the more
negative is $K$, the faster is the collapse.

\subsection{Algebraic Decay of the Frequency}

The frequency
\begin{equation}
\label{y1} \omega = \frac{\sqrt{3}}{(1 + t^2)}
\end{equation}
illustrate the relevant case of a trapping potential that decays at an algebraic rate.
The numerical coefficients in (\ref{y1}) are chosen so as to obtain simple expressions
in the following. For (\ref{y1}), the solution for Pinney's equation with the
prescribed initial condition is
\begin{equation}
\label{y2} \rho = \left(\frac{t^4 - t^2 + 1}{t^2 + 1}\right)^{1/2} \,.
\end{equation}
The final expression for $r$ as a function of $t$ which follows from (\ref{x4}) is
given by
\begin{equation}
r^2 = \frac{(1-t^2)^2 + 2Kt^2}{t^2 + 1} \,,
\end{equation}
showing that $r \rightarrow t$ asymptotically. Using (\ref{x3}), the critical time for
collapse when $K \leq 0$ is obtained as
\begin{equation}
t^* = \sqrt{1 + |K|/2} - \sqrt{|K|/2} \,,
\end{equation}
again a decaying function of $|K|$. Other kinds of decaying traps are also amenable to
exact calculations, as for the case of an exponential decay, which can be treated in
terms of Bessel functions.

\subsection{Sudden Jump in the Frequency}

As a final illustrative example, consider the case of a sudden jump in the frequency
at some time $\tau > 0$,
\begin{equation}
\omega = \cases {0 \,, \quad t < \tau \,;\cr 1 \,, \quad t > \tau \,. \cr}
\end{equation}
For this choice, the solution for Pinney's equation with the prescribed initial
condition is
\begin{equation}
\rho = \cases {(1 + t^2)^{1/2} \,, \quad t < \tau \, ; \cr (\cos(t - \tau)^2 +
(\tau\cos(t - \tau) + \sin(t - \tau))^2)^{1/2} \,, \quad t > \tau \,. \cr}
\end{equation}
This solution is continuous and has continuous first derivative at $t = \tau$. Using
(\ref{x4}), we obtain
\begin{equation}
r = \cases {(1 + 2K\,t^2)^{1/2} \,, \quad t < \tau \, ; \cr (\cos(t - \tau)^2 +
2K\,(\tau\cos(t - \tau) + \sin(t - \tau))^2)^{1/2} \,, \quad t > \tau \cr}
\end{equation}
for the time-evolution of $r$, which becomes to be oscillatory at $t = \tau$ if it has
not collapsed before. Indeed, by (\ref{x3}) the critical time for non-positive $K$ can
attain two expressions according to the values of $\tau$ and $|K|$,
\begin{equation}
t^* = \cases {1/\sqrt{2|K|} \,, \quad \tau > 1/\sqrt{2|K|} \,; \cr \tau +
\arctan(1/\sqrt{2|K|} - \tau) \,, \quad \tau < 1/\sqrt{2|K|} \,. \cr}
\end{equation}
No matter the values of the parameters $\tau$ and $|K|$, the critical time is a
decaying function of $|K|$ as it should be.

\section{Conclusion}

The main conclusion of this work is the inequality (\ref{eqqq24}), which is a
necessary and sufficient condition for collapse. This condition shows that the
decisive quantity for collapse is the invariant $K$, the Ermakov invariant apart from
an additive constant, and not the coupling constant $g$. Sufficient angular moment can
prevent collapse, even in the case of highly attractive condensates. This conclusion
was guided by the description of the Bose-Einstein condensate in terms of a completely
integrable Ermakov system. We remark that the exact solution shown in Section 3 is the
exact solution for a variational approach based on a Gaussian {\it ansatz}. This gives
a true exact solution for the Gross-Pitaevskii equation only in the Thomas-Fermi
limit, when the first terms at the right-hand sides of equations (\ref{g1}-\ref{g2})
are disregarded. Nevertheless, the exact solution is an improvement over isotropic
solutions (with zero angular moment) or self-similar solutions \cite{rybin2}.
Furthermore, our techniques allow possible time-dependence of the trap.

A na\-tu\-ral ques\-tion is about the ex\-ten\-sion of this work to
three-di\-men\-sio\-nal Bose-Einstein condensates. In this case, the Gaussian {\it
ansatz} for the Gross-Pitaevskii action does not lead to Ermakov systems, but to
Ermakov systems perturbed by terms directly proportional to the coupling constant $g$.
This suggests the use of perturbation techniques for small $g$. This direction is now
being pursued in a work in progress.

\vskip 1cm \noindent{\bf Acknowledgements}\\ We are grateful to A. C. Olinto for
valuable comments and suggestions. This work has been supported by the Brazilian
agency Conselho Nacional de Desenvolvimento Cient\'{\i}fico e Tecnol\'ogico (CNPq).

\end{document}